\documentclass[11pt]{article}
\usepackage{graphicx}

\begin{document}

\begin{center}
  \textbf{\Large Multidimensional HBT correlations\\[6pt]
    in $\mathbf{p\bar p}$ collisions at $\mathbf{\sqrt{s} = 630}$ GeV}
  \footnote{Contribution to the \textit{Workshop on Particle
      Correlations and Femtoscopy}, Krom\v e\v r\'i\v z, Czech
    Republic, 15--17 August 2005. To appear in: \textit{Proc.\ 35th
      International Symposium on Multiparticle Dynamics}, Krom\v e\v
    r\'i\v z, 9--15 August 2005, American Institute of Physics
    (2006).}
  \\[12pt]
  H.C.~Eggers$^{ a}$, B.~Buschbeck$^{ b}$ and F.J.~October$^{ c,a}$\\[12pt]

  \textit{$^a$ Department of Physics, University of Stellenbosch,
    7602 Stellenbosch, South Africa}\\
  \textit{$^b$ Institut f\"ur Hochenergiephysik,
           Nikolsdorfergasse 18, A--1050 Vienna, Austria}\\
  \textit{$^c$ Institute for Maritime Technology, 
           7995 Simonstown, South Africa}\\
\end{center}
\begin{abstract}
  \noindent
  We analyse second moments $R_2$ of like-sign pion pairs in the
  two-dimensional $(q_L,q_T)$ and three-dimensional $(q_O,q_S,q_L)$
  decompositions of the three-momentum difference. Conventional fit
  parametrisations such as gaussian, exponential, power-law and
  Edgeworth fail miserably, while more elaborate ones such as L\'evy
  do well but fail to yield a unique best-fit solution. A
  two-component model using a hard cut to separate small- and
  large-scale parts appears possible but not compelling. In all cases,
  the data exhibits a strong and hitherto unexplained peak at small
  momentum differences which exceeds current fits.
\end{abstract}
\begin{quote}
Classification: 13.85.Hd, 13.87.Fh, 13.85.-t, 25.75.Gz
\end{quote}


The UA1 experiment, having completed data-taking at the CERN SPS in
the late eighties, continues to be relevant and interesting. The
current concentrated effort at RHIC to quantify and understand
ultrarelativistic nuclear collisions relies extensively on comparisons
with baseline scenarios constructed from the corresponding ``trivial''
hadron-hadron sample. Current experimental energies of $200$ AGeV at
RHIC are still below those available to UA1 by a factor three, so that
UA1 results may also provide a window on possible energy dependencies
of current investigations.
\\

In this contribution, we provide preliminary results on HBT analysis
mainly in terms of the two-dimensional decomposition of the
three-momentum difference, defining in the usual way $q_L =
|\mathbf{q}_L| = |(\mathbf{q}\mathbf{\cdot}\hat z)\, \hat z|$, with
$\hat z$ the beam direction, and $q_T = |\mathbf{q} - \mathbf{q}_L|$.
Brief reference is also made to the three-dimensional Bertsch-Pratt
case, with similar results and issues arising.
\\


Like-sign (LS) pion pairs from approximately 2.45 million minimum-bias
events measured by the UA1 central detector were analysed. This
represents a twofold increase in statistics over Ref.~\cite{Bus00a}
and a 15-fold increase compared to earlier UA1 HBT analyses
\cite{UA1-89b,UA1-92a,Egg97a}.  Standard cuts \cite{Bus00a} were
applied, including single-track cuts $p_\bot \geq 0.15$~GeV/c, $| y |
\leq 3$ and, to avoid acceptance problems, $45^\circ \leq |\phi| \leq
135^\circ$. The sample contains an estimated 15\% contamination of
charged kaons.
\\

The most important among the standard pair cuts is the ``ghost cut''
which eliminates spurious ``split track'' LS pairs within a narrow
cone but many real LS pairs also. A correction factor compensating for
this was determined by passing unlike-sign (US) pairs through the same
ghost-cut algorithm. Such correction factors were determined for each
$(q_L,q_T)$ bin\footnote{The $(q_L,q_T)$ bins for ghost corrections
  use $q_L$ measured in the detector rest system ($p\bar p$ CMS),
  while all HBT quantities are measured in the LCMS.} and
charged-multiplicity subsample; for some $(q_L,q_T)$ bins, this
correction factor ranges up to 1.7 or even 1.9 for low-multiplicity
subsamples.
\\

We corrected for Coulomb repulsion by parametrising the Bowler Coulomb
correction in the invariant momentum difference $Q =
\sqrt{-(p_1-p_2)^2}$ \cite{Bow91a} with an exponentially damped Gamov
factor $G(Q)$ \cite{Bri95a}, with a best-fit value $Q_{\rm eff} =
0.173 \pm 0.001$ GeV/c,
\begin{equation}\label{dsb}
F_{\rm coul}(Q) = 1 +  \left[ G(Q) - 1 \right] 
                   \, \exp(-Q/Q_{\rm eff}) \,.
\end{equation}
The reference sample was formed by randomly combining LS tracks taken
from pools of events in the same subsample of event charged
multiplicity $N$ as the sibling event currently being analysed. Note
that the reference for fixed-multiplicity subsamples is not the
poisson distribution but the multinomial, whose second moment is
$\rho_2^{\rm mult} (\mathbf{q}\,|\,N) = (1-N^{-1})
\rho_1{\otimes}\rho_1(\mathbf{q}\,|\,N)$, so that the appropriate
normalised moment is \cite{Bus00a,Lip96a}
\begin{equation}\label{dsc}
R_2(\mathbf{q}) = 
\frac
{\sum_N P_N\, \rho_2^{\rm LS}(\mathbf{q}\,|\,N)}
{\sum_N P_N\, (1-N^{-1})\,\rho_1{\otimes}\rho_1^{\rm LS}(\mathbf{q}\,|\,N)}
\,.
\end{equation}
\par\vspace*{-11mm}
\begin{center}
\includegraphics[width=170mm,viewport=40 0 595 255]
       {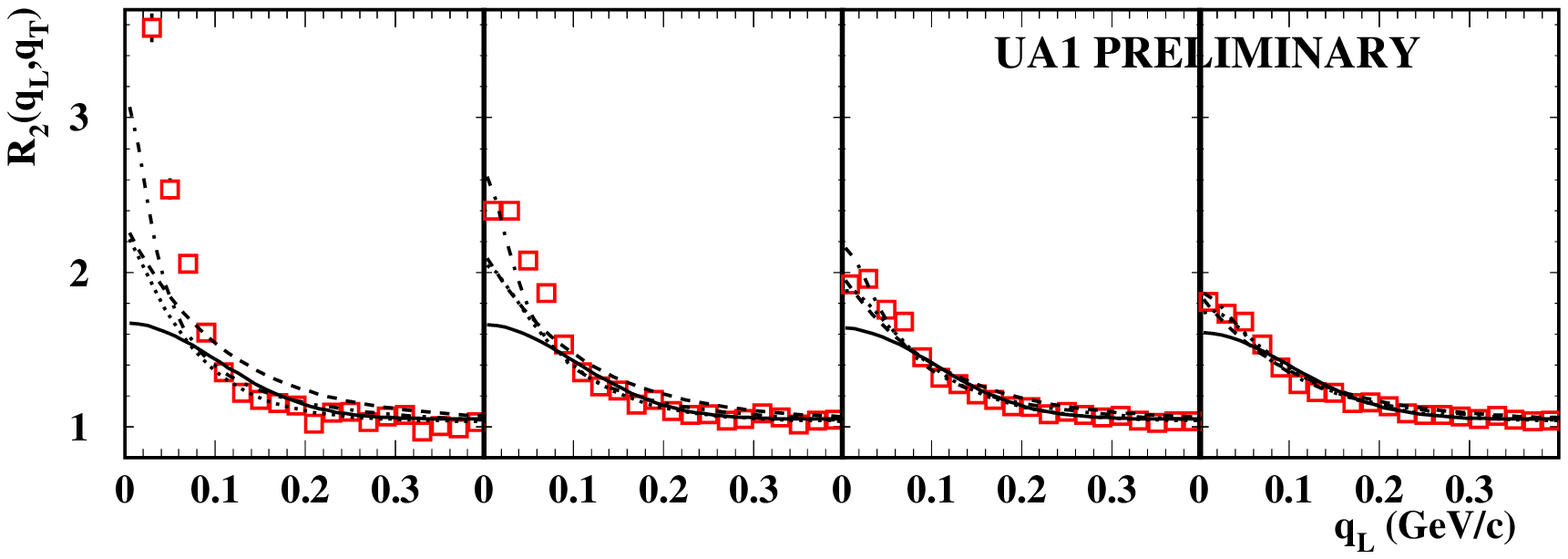}\\[-19mm]
\includegraphics[width=170mm,viewport=40 0 595 255]
       {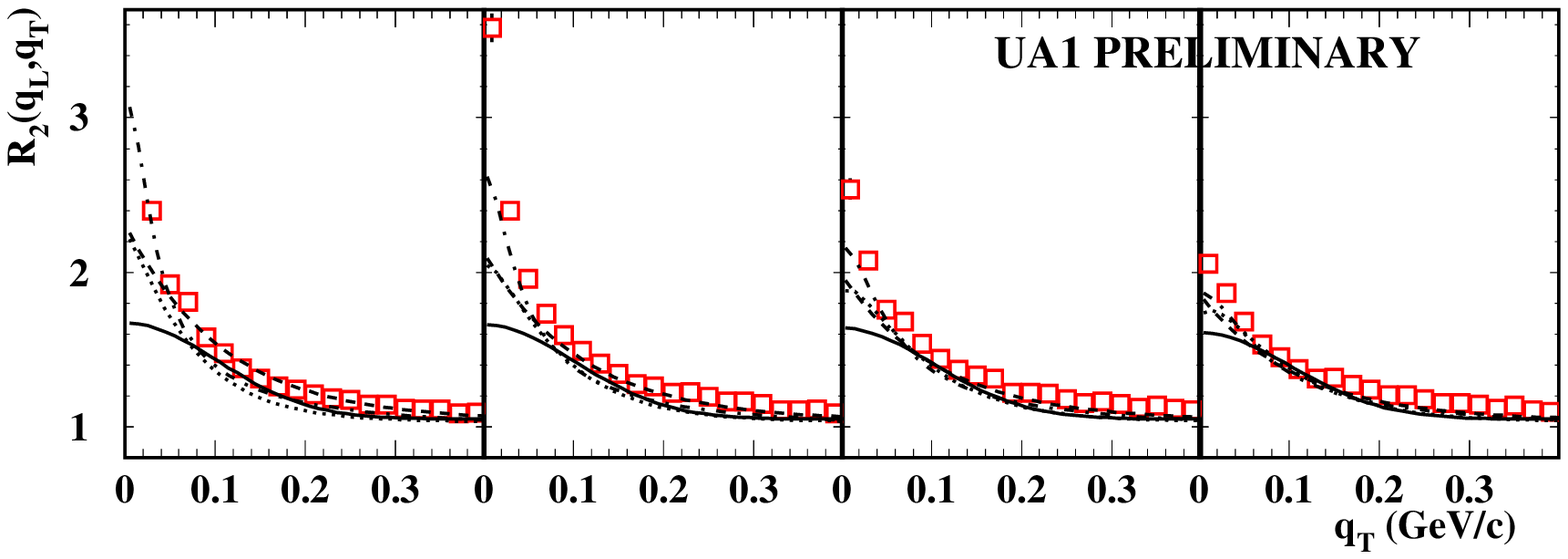}\\
\end{center}
\par\vspace*{-12mm}
\begin{quote} {\small \textbf{FIGURE 1.} Upper panels: $R_2(q_L,q_T)$
    data and best fits, shown for the first slices (left to right)
    $q_T = $ 0.00--0.02, 0.02--0.04, 0.04--0.06 and 0.06--0.08 GeV/c.
    Lower panels: $R_2(q_L,q_T)$ and the same fits for corresponding
    fixed-$q_L$ slices.  Solid lines: Gauss/Edgeworth fits; dashed:
    exponential; dotted: exponential with cross term; dash-dotted:
    L\'evy.  The bin $(q_L,q_T) < (0.02,0.02)$ GeV/c is omitted from
    all fits and plots.}
\end{quote}
\ \\[1mm]

The ghost- and coulomb-corrected normalised moment $R_2(q_L,q_T)$ is
shown in Fig.~1, together with fits to parametrisations %
$R_2 = \gamma[ 1 + \lambda \left| S_{12}\right|^2 ]$, with %
$\left| S_{12} \right|^2$ parametrised respectively as %
$\exp(-R_L^2 q_L^2 - R_T^2 q_T^2 - 2 R_{LT} q_L q_T)$ %
(gauss with cross term), %
$\exp(-R_L q_L - R_T q_T)$ (exponential), %
$\exp(-R_L q_L - R_T q_T - 2R_{LT} \sqrt{q_L q_T})$ %
(exponential with cross term) and %
$R_2 = \gamma \left[ 1 + \left(R_L q_L \right)^{-\alpha_L} \left(R_T
    q_T \right)^{-\alpha_T} \right]$ %
(product power law). Note that all results shown are preliminary.
\\

It is immediately apparent that none of these fits reproduces the
data, with $\chi^2/\mathrm{NDF}$ ranging from 3 to 9. A
parametrisation based on an Edgeworth expansion \cite{Cso93a,STAR05a},
\begin{equation}\label{rse}
\left| S_{12} \right|^2 = \exp(-R_L^2 q_L^2 - R_T^2 q_T^2)
\prod_{d=L,T}\left[ 
1 + \kappa_{4,d}\; H_4(\sqrt{2}\,R_d\,q_d)/24 \right],
\end{equation}
(with $\kappa_{4,d}$ the fourth-order cumulant in $q_d$ and $H_4$ the
corresponding hermite polynomial) fares no better as best-fit values
for $\kappa_{4,d}$ in both directions turn out to be negligible. We
omit the third-order terms in $\kappa_{3,L}$ and $\kappa_{3,T}$ as
they are antisymmetric in $q_d$.
\begin{center}
\ \\[14mm]
\includegraphics[width=130mm,viewport=60 0 590 369]
       {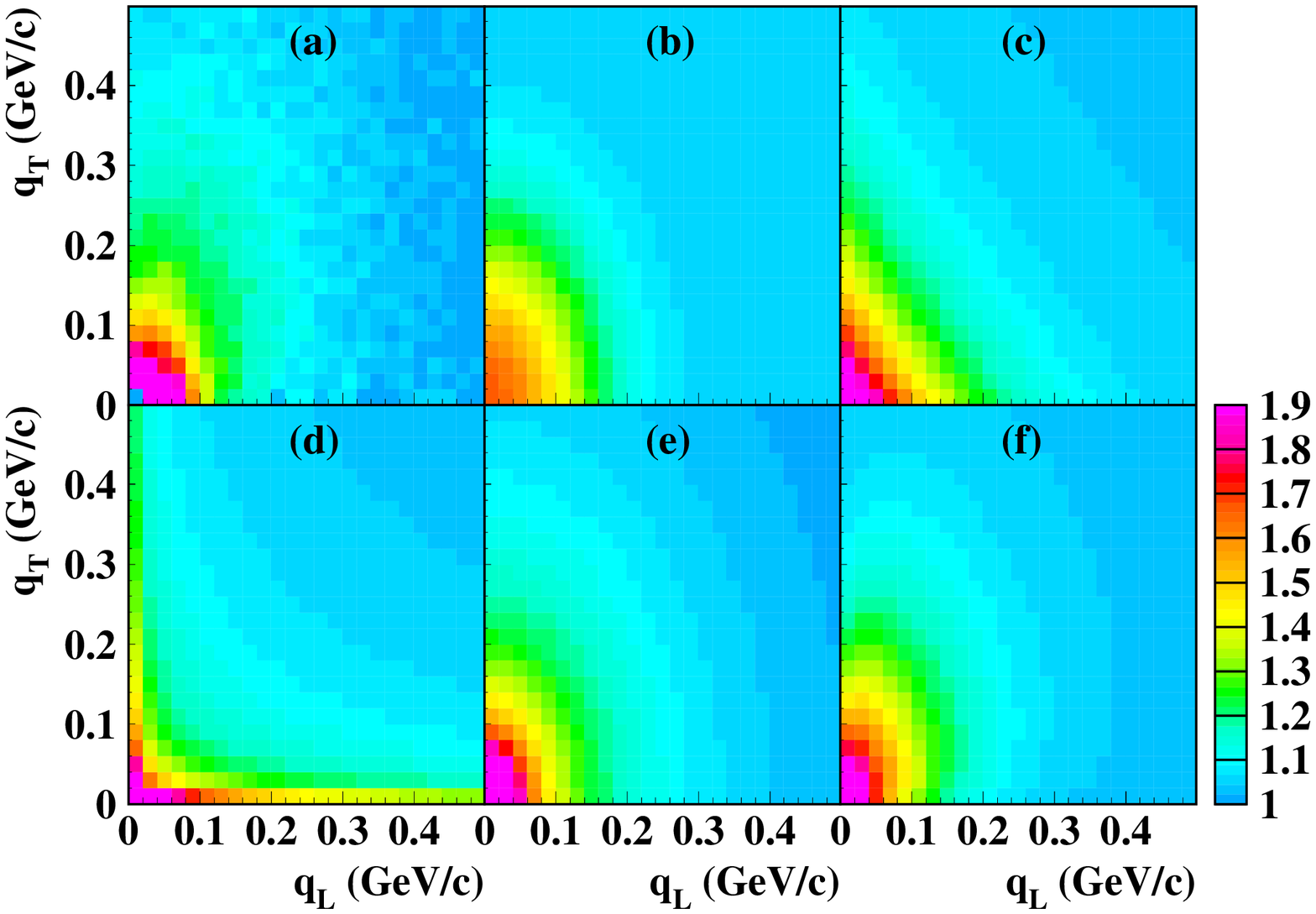}\\
\end{center}
\par\vspace*{-5mm}\par
\begin{quote} {\small \textbf{FIGURE 2:} Comparing the shapes of
    $R_2(q_L,q_T)$ at intermediate scales. Panels show (a)) UA1 data,
    (b) gauss, (c) exponential, (d) power-law, (e) L\'evy, (f)
    exponential with cross term fits. The gauss-type fits have the
    right shape, but end up far below the data peak at small
    $(q_L,q_T)$.  While the simple exponential and product power-law
    may approximate the peak reasonably, both fail miserably when
    shapes at intermediate scales are considered. L\'evy does well but
    no unique best fit can be found. Shape-wise, the exponential with
    cross term (f) appears to come out on top.  All plots are
    truncated at $R_2 \leq 1.9$ in order to bring out structure at
    intermediate scales.  }
 \end{quote}
\ \\[-3mm]

\noindent
A L\'evy-based parametrisation  \cite{Cso04a},
\begin{equation}\label{rsf}
\left| S_{12} \right|^2 = \exp(-R_L^2 q_L^2 - R_T^2 q_T^2)^{\alpha/2},
\end{equation}
\noindent
yields better results in reproducing the strong peak observed in the
data; however, four of the five fit parameters, viz.\ $\lambda$,
$R_L$, $R_T$ and $\alpha$, are strongly correlated so that no unique
best fit can be achieved. One example of many equivalent ``best'' fits
is shown in Fig.~1. Omitting a second small-$(q_L,q_T)$ bin from the
L\'evy fit renders the fit even more unstable (the innermost bin
$(q_L,q_T) < (0.02,0.02)$ GeV/c is excluded from all analysis as a
matter of course).  This is hardly surprising, as the four
abovementioned parameters collectively depend strongly on the exact
shape of the peak in the very small $(q_L,q_T)$ region --- the very
region that experimental measurement struggles to resolve.
\\

Interesting, nonetheless, is the observation that fits fail for
different reasons: As shown in Fig.~2, the gauss and Edgeworth fits
reproduce the shape of $R_2$ at intermediate $(q_L,q_T)$ rather well
but lacks the strong peak exhibited in the data,\footnote{Edgeworth
  fits are indistinguishable from the normal gaussian fits throughout
  this analysis.}  while the exponential and product power-law
parametrisations are more peaked but fail to describe the shape at
intermediate $(q_L,q_T)$.  Judging by shape alone, the exponential
with a $2 R_{LT}\sqrt{q_L\, q_T}$ cross term does best (Fig.~2(f)).
\\

Either way, it is clear that the superiority of the power-law fit to
the one-dimensional invariant four-momentum moment $R_2(Q)$ over Gauss
and exponential fits \cite{Egg97a} is not repeated in the
$R_2(q_L,q_T)$ case due to the hyperbolic shape of our ``product
power-law'' parametrisation\footnote{Other forms, such as a ``sum
  power law'' $R_2 = \gamma\left[1 + (R_L\, q_L)^{-\alpha_L} + (R_T\,
    q_T)^{-\alpha_T} \right]$ remain to be tested.} compared to the
elliptic form of the $R_2(q_L,q_T)$ data shown in Fig.~2(a). The
analysis of shapes at intermediate scales appears to provide valuable
information.  Shape analysis in the form of Refs.~\cite{Dan05a,
  Lis05a} may help to quantify these qualitative observations.
\\

The failure of conventional parametrisations to reproduce the elusive
peak at small $(q_L,q_T)$ suggests that there may be two scales in the
system.  Borrowing the (strictly speaking inappropriate) terms
``core'' and ``halo'' from the literature \cite{Cso94a}, we try a
two-scale ``butcher's model'', similar to simpler precursors in Refs.\
\cite{Led79a,NA22-93a},
\begin{equation}\label{rsg}
R_2(q_L,q_T) = \gamma 
\left[ 1 
   + \lambda_C \exp(-R_{LC}^2 q_L^2 - R_{TC}^2 q_T^2 )
   + \lambda_H \exp(-R_{LH}^2 q_L^2 - R_{TH}^2 q_T^2 )
\right],
\end{equation}
and proceed as follows: First, we fit only bins with momentum
differences larger than a hard cutoff, $(q_L,q_T) > (q_{\rm
  cut},q_{\rm cut})$, to the core gaussian. The resulting best-fit is
subtracted from all data.  The remaining halo ``data'' with $(q_L,q_T)
\leq (q_{\rm cut},q_{\rm cut})$ is then fit with the halo gaussian,
and the resulting ``core'' and ``halo'' best fits combined. Data and
fit histograms at various steps of this procedure are shown in Fig.~3.
\\[4mm]

\par\hspace*{20mm}
\includegraphics[width=130mm,viewport=80 0 567 567]
  {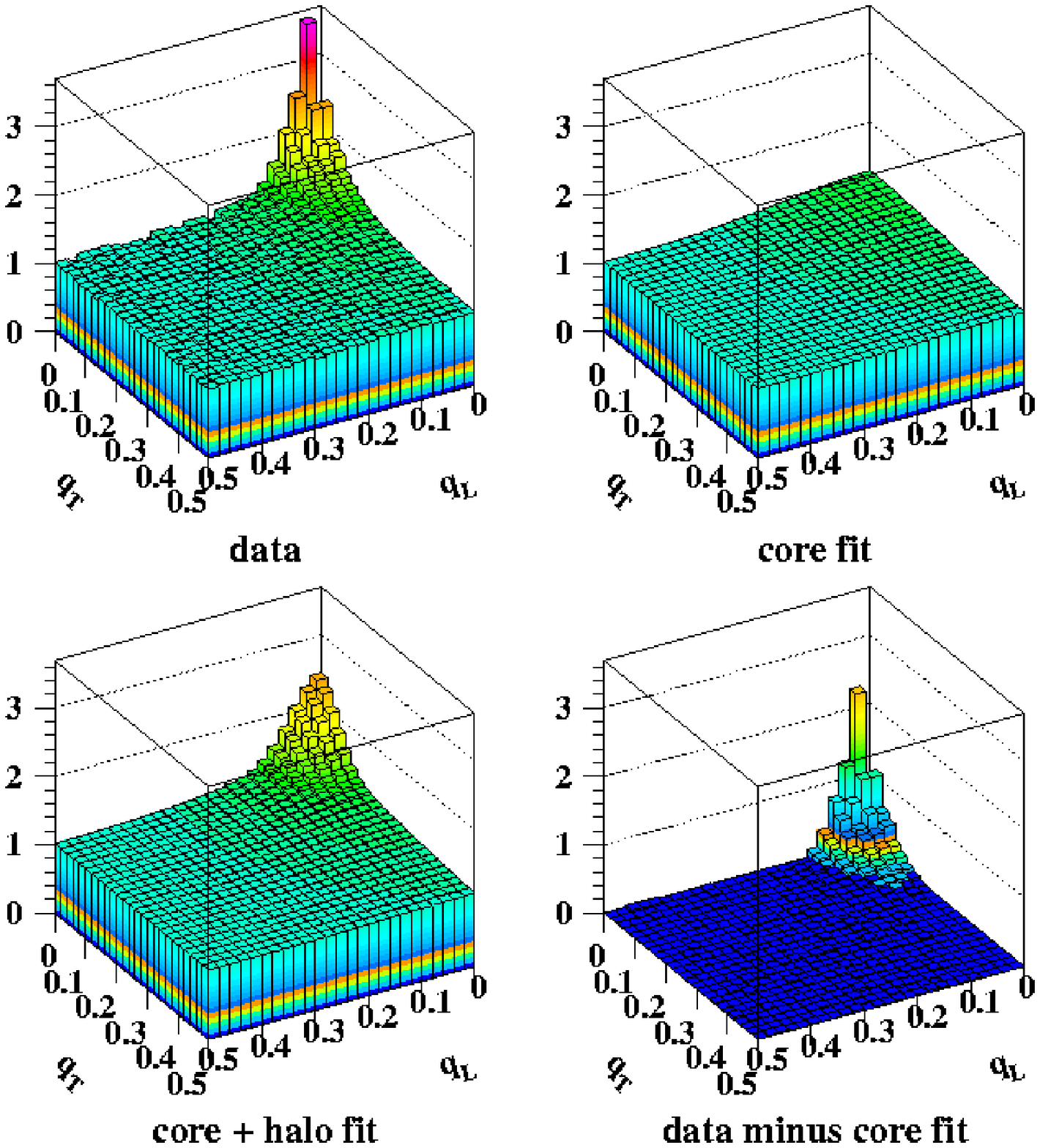}
\begin{quote} {\small \textbf{FIGURE 3:} Two-scale but\-cher's model.
    Of the $R_2(q_L,q_T)$ data (left upper panel) only bins with
    $(q_L, q_T) > (0.16,0.16)$ GeV/c are fit to the ``core'' gaussian
    of Eq.~(\ref{rsg}) as shown in the right upper panel.  The result
    is subtracted from the data to reveal the ``halo'' (right lower)
    which is then fit to the halo part of Eq.~(\ref{rsg}).  The
    combined core-halo fit is shown in the left lower panel.}
\end{quote}

\par\hspace*{10mm}
\includegraphics[width=150mm,viewport=0 0 510 390] 
     {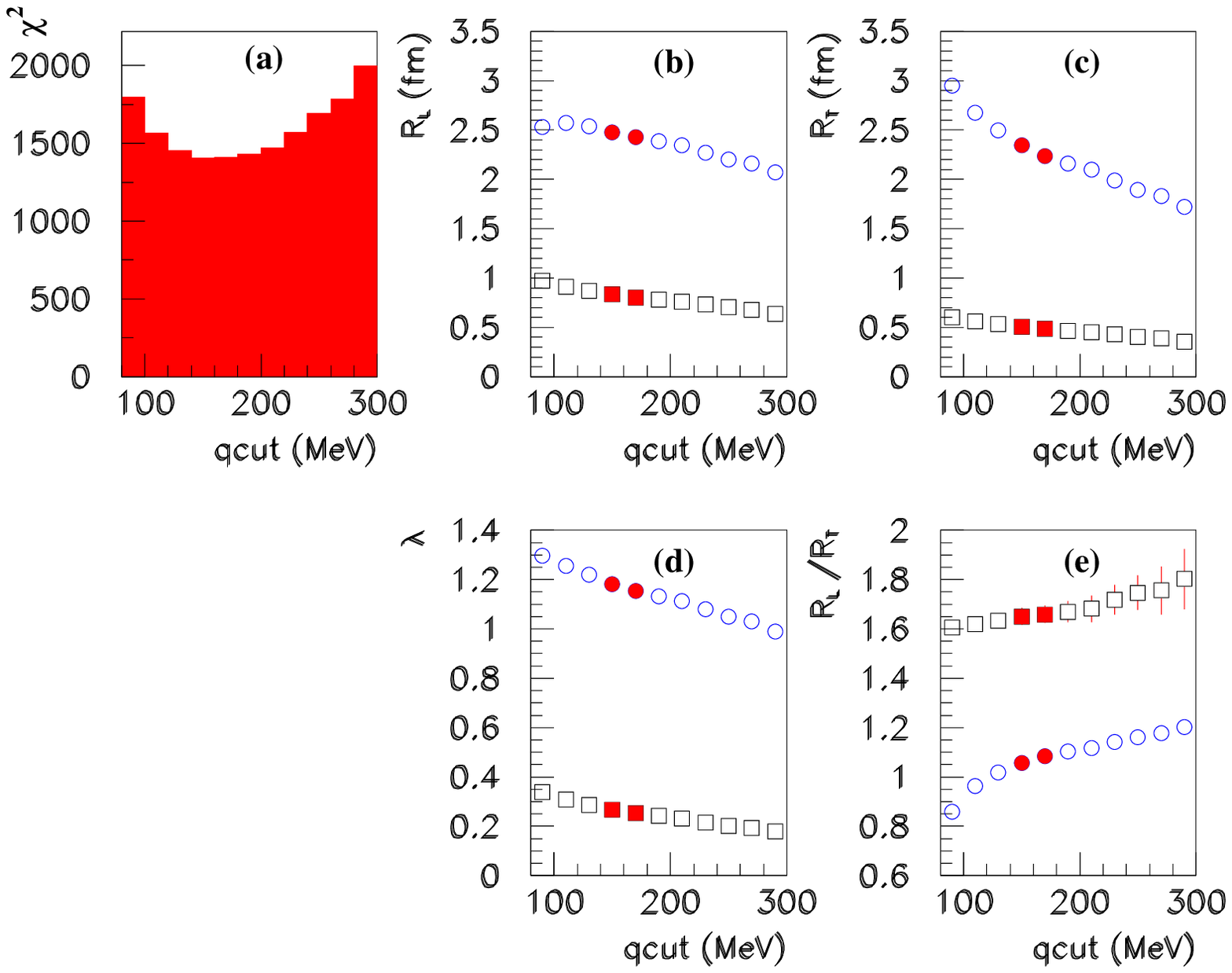}
\\[-10mm]
\begin{quote}
  {\small \textbf{FIGURE 4:} Dependence on $q_{\rm cut}$. (a) Total
    $\chi^2$ for both core and halo fits as a function of $q_{\rm
      cut}$.  The two lowest $\chi^2$ values correspond to $q_{\rm
      cut} = 0.16\ \mbox{and\ }0.18$ GeV/c, giving
    $\chi^2/\mathrm{NDF} = 2.28$.  Panels (b)--(e): Dependence of
    best-fit parameter values on $q_{\rm cut}$.  Squares and circles
    represent core and halo fit parameters respectively. A clear
    separation of scales is observed as assumed in the model.  The
    halo continues to have a chaoticity parameter $\lambda_H$ that
    exceeds its theoretical limit of 1.}
\end{quote}
\ \\[-3mm]

\noindent
In Fig.~4, the dependence of the two-scale model on $q_{\rm cut}$ is
tested. As shown in Fig.~4(a), the joint $\chi^2$ for both fits as a
function of $q_{\rm cut}$ is found to have a well-defined minimum for
$q_{\rm cut} =$ 160--180 MeV/c. The best estimates for $R_L$, $R_T$,
$\lambda$ and $R_L/R_T$ are the two filled points in Fig.~4(b)--(e)
corresponding to the two best $q_{\rm cut}$ values.  Averaging these
two numbers, we estimate $R_{LC} = 0.82\pm 0.02$ fm, $R_{TC} = 0.49\pm
0.02$ fm, $R_{LH} = 2.45\pm 0.03$ fm, $R_{TH} = 2.29\pm 0.05$ fm,
$\lambda_C = 0.26 \pm 0.01$ and $\lambda_H = 1.17 \pm 0.02$,
signalling a prolate ``core'' and a roughly spherical ``halo''. The
clear separation between the sizes of the core and halo radii
\textit{a posteriori} support the assumption of the presence of two
scales, i.e.\ signal that the two-scale model is consistent.  We note
that $\lambda_H$ continues to exceed the theoretical limit of 1.00,
albeit not as strongly as the huge intercept $R_2(0,0) > 2.7$ seen in
the data itself. \textit{The joint best $\chi^2/{\rm NDF} = 2.28$ for
  the two-scale model is still rather large, however, so that all
  numbers and conclusions should be treated with caution.}
\\

Turning briefly to the more common three-dimensional Bertsch-Pratt
representation \cite{Ber88a}, we find that the data once again has a
strong peak at small $(q_O,q_S,q_L)$, and that the simple gaussian
parametrisation fails completely. In Fig.~5, we show the result of
fitting the L\'evy and Edgeworth parametrisations,
\begin{eqnarray} \label{rsk}
\left| S_{12}\right|^2
 &=& \exp(-R_O^2 q_O^2 - R_S^2 q_S^2 - R_L^2 q_L^2)^{\alpha/2},
\\
\left| S_{12}\right|^2
&=& \exp(-R_O^2 q_O^2 - R_S^2 q_S^2 - R_L^2 q_L^2)
\prod_{d=O,S,L}\left[ 
   1 + \kappa_{4,d}\; H_4(\sqrt{2}\,R_d\,q_d)/24 \right],
\end{eqnarray}
to $R_2(q_O,q_S,q_L)$, with slices plotted along the three axes of the
three-dimensional space. Even these parametrisations appear not to
describe the data well for small $|\mathbf{q}|$, and in this
threedimensional case the discrepancies are spread more widely than in
the $(q_L,q_T)$ case. Also, while the L\'evy fit does appear to do
reproduce the data to a reasonable degree, it again suffers from
strong inherent correlations between the parameters $\lambda, R_O,
R_S, R_L$ and $\alpha$. Such correlation again implies that no unique
minimum for the $\chi^2$ exists and thereby no unique best-fit values
for its parameters either.
\\[15mm]

\par\hspace*{12mm}
\includegraphics[width=140mm,viewport=80 0 520 184] 
       {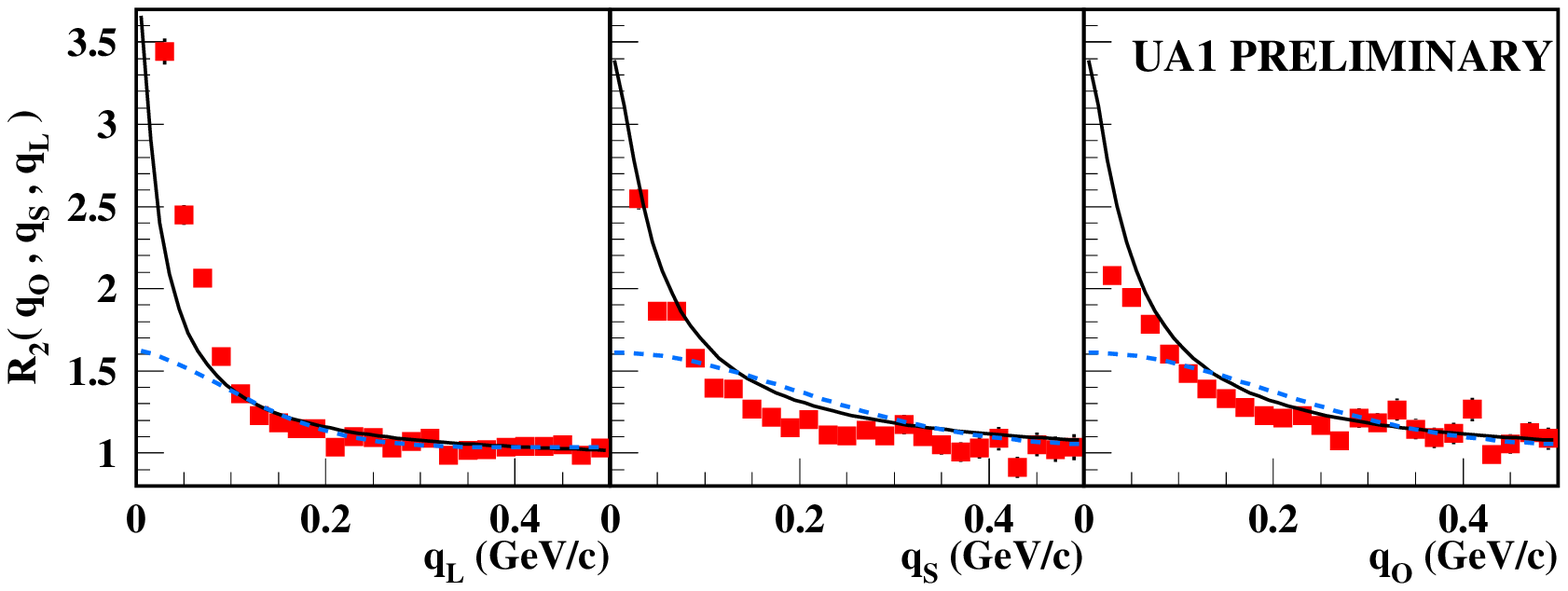}
\par\vspace*{-12mm}\par
  \begin{quote}
    {\small \textbf{FIGURE 5:} Second moment $R_2(q_O,q_S,q_L)$ of
      three-dimensional momentum difference decomposition, plotted
      along the three axes $q_d$, with the other two variables
      $(q_e,q_f) < (0.02,0.02)$ GeV/c.  Solid line: L\'evy fit; dashed
      line: Edgeworth. }
  \end{quote}
  \ \\[-2mm]

\noindent
It is, of course, always possible to try even more elaborate
parametrisations, for example using separate $\alpha$-exponents for
each of the $(q_O,q_S,q_L)$ directions.  While a better fit might
theoretically be achieved, this will invariably come at the price of
even more highly correlated parameters. The data cannot distinguish
between many combinations of ``best'' values for such parameters.
\\

We note that, even within the arguably artificial method of the
two-scale model, the strong peak in the data is not reproduced,
resulting in the quoted low confidence level. Nevertheless, it may
provide a useful hint that ``something else is going on'', be it the
influence of jets, clustering effects or some other unknown factor. 
\\

We stress that is is unlikely that the strong peak seen in our data at
small momentum differences, over and above simple gaussian or
exponential parametrisations, is due to bias. First, the peak
persists even without the Coulomb or split-track corrections.  Second,
we believe that the same peak may well be responsible for the
power-law form and higher-order effects seen in our earlier work with
one-dimensional distributions \cite{Egg97a}.  Third, the presence of
unidentified kaons and protons in the sample imply that the real peak
should exceed the one shown here, so that the present results may be
lower limits.  It should also be noted that a number of other hadronic
experiments have previously seen significant deviation from gaussian
behaviour, especially at small $|\mathbf{q}|$ \cite{AFS-83a,NA22-95a}. The
challenge is clearly now to find a convincing physical cause and
explanation.
\\


\textbf{Acknowledgements:}
  This work was supported in part by the National Research Foundation
  of South Africa. BB thanks the University of Stellenbosch for
  kind hospitality.


\end{document}